\documentclass{osa-article}

\journal{oe}
\usepackage{color}
\usepackage{ulem}
\usepackage{booktabs}
\articletype{Research Article}

\begin{document}
\title{Polarization design for ground-to-satellite quantum entanglement distribution}

\author{Xuan Han,\authormark{1,2,4} Hai-Lin Yong,\authormark{1,2,4} Ping Xu,\authormark{1,2} Kui-Xing Yang,\authormark{1,2} Shuang-Lin Li,\authormark{1,2} Wei-Yang Wang,\authormark{1,2} Hua-Jian Xue,\authormark{1,2} Feng-Zhi Li,\authormark{1,2} Ji-Gang Ren,\authormark{1,2,3} Cheng-Zhi Peng,\authormark{1,2} and Jian-Wei Pan\authormark{1,2}}

\address{\authormark{1}Department of Modern Physics and Hefei National Laboratory for Physical Sciences at the Microscale, University of Science and Technology of China, Hefei 230026, China\\
\authormark{2} Chinese Academy of Sciences (CAS) Center for Excellence and Synergetic Innovation Center in Quantum Information and Quantum Physics, University of Science and Technology of China, Shanghai 201315, China}

\email{\authormark{3}jgren@ustc.edu.cn}
\email{\authormark{4}These authors contributed equally to this work.} 



\begin{abstract}
High-fidelity transmission of polarization encoded qubits plays a key role in long distance quantum communication. By establishing the channel between ground and satellite, the communication distance can even exceed thousands of kilometers. Aim to achieve the efficient uplink quantum communication, here we describe a high-fidelity polarization design of a transmitting antenna with an average polarization extinction ratio of 887:1 by a local test. We also implement a feasible polarization-compensation scheme for satellite motions with a fidelity exceeding 0.995. Based on these works, we demonstrate the ground-to-satellite entanglment distribution with a violation of Bell inequality by 2.312$\pm$0.096, which is well above the classic limit 2.
\end{abstract}



\section{Introduction}
Entanglement is a physical phenomenon that has puzzled many scientists in the last century\cite{EPR_35,Schrodinger:Entanglement:1935}. In 1965, Bell proposed a mathematical model later named after him to test whether entanglement is caused by local hidden variables\cite{Bell_Ineq_64}. The violations of Bell inequality have been observed in many experiments\cite{Freedman.Bell.72, Aspect:Belltest:1982,Hensen:NVLoophole:2015,giustina:2015:loophole-free}, indicating that the local-hidden-variable theorem does not apply in these cases. At the same time, the distances of outdoor entanglement distribution have extended from hundreds of meters to thousands of kilometers\cite{Aspelmeyer:entangle:2003,Peng:13km:2005,Yin:Teleportation:2012,Yin:SatEPR:2017}, providing tests of quantum nonlocality on a large scale. Practically, long-distance entanglement distribution also plays an important role in quantum communications such as quantum teleportation \cite{bennett:1993:tele,Bouwmeester97:Teleportation,Ren:SatTele:2017} and entanglement-based quantum key distribution\cite{Bennett:BBM92:1992,Yin:2017:satellite_QKD}.

Polarization encoding of photons is always the first choice for long-distance free space quantum communications\cite{Yin:Teleportation:2012,XSMa:Teleportation:2012}, including satellited-based entanglement distribution\cite{Yin:SatEPR:2017}. However, the polarization states will be affected by optical elements and the changing reference frame due to satellite motions\cite{bonato:2006:sat_motion_polarization,Bonato:2007:influence,Zhang:Polarization:2014}. Therefore, polarization maintenance and compensation of photons is one of the essential technology. Apparently, both downlink and uplink transmissions are necessary to construct a satellite-based global quantum communication. Many downlink measurements of polarization have been carried out by different research groups\cite{Toyoshima:polarization:2009,Yin:single:2013,Vallone:Single:2015,Japan:Satellite:2017}. Recently, some satellite-based downlink quantum communications\cite{Yin:2017:satellite_QKD,Liao:SatQKD:2017} have taken an important step toward global quantum communications. While, few uplink experiments\cite{Ren:SatTele:2017} have been reported due to higher link loss\cite{bourgoin:2013:comprehensive}. 

In this paper, we introduce a polarization design for a transmitting antenna at the Ngari optical ground station for the quantum science satellite (Micius). With tailored multilayer coatings, we obtain polarization extinction ratios (PERs) greater than 400:1, with an average PER of 887:1. Using a periscope structural design of the transmitting antenna, we implement a polarization-compensation scheme for single-mode fibers (SMFs) and satellite motions. As a result, the uplink fidelity of polarization is approximately 0.995. Finally, we also demonstrate ground-to-satellite entanglement distribution for the first time. 

This paper is organized as follows. In section 2, discussion is focused on the polarization design of a transmitting antenna and local test of the transmitting antenna. In section 3, the polarization compensation for the fiber and satellite motion is implemented and the ground-to-satellite polarization fidelities are measured. In section 4, an uplink entanglement distribution experiment is demonstrated. In section 5, a summary and conclusion is provided.

\section{Polarization-maintenance of the transmitting system}

  Using high-precision acquisition, pointing and tracking (APT) technology\cite{Han:2018:point-ahead}, we can build an uplink channel between the ground and Micius. There are three major elements that will affect the polarization: the transmitting system on the ground, the receiving system on the satellite \cite{Wujincai:2017:polarization} and the channel. The transmitting and receiving systems must be well designed to maintain the polarization. The optical design of the transmitting antenna is described below. As the phase variation caused by atmosphere are less than $5.22\times 10^{-9}~\mu rad$\cite{hohn:1969:depolarization}, so that it can be ignored in our experiment. The channel influence is mainly caused by satellite motions and the SMFs. The polarization compensation against satellite motions and SMFs will be discussed in the next section. 

\subsection{Influence of reflective optical systems}
   As we know, the reflectances of s-polarization and p-polarization light are different, which follow the Fresnel equations \cite{Born:1999:principles}:
   \begin{eqnarray}
   r_s=\dfrac{n_0\cos\theta_i-n\cos\theta_t}{n_0\cos\theta_i+n\cos\theta_t}\\
    r_p=\dfrac{n\cos\theta_i-n_0\cos\theta_t}{n\cos\theta_i+n_0\cos\theta_t}
   \end{eqnarray}
  Where $r_s$ and $r_p$ are the reflectances for s-polarization and p-polarization light, respectively (the power reflection coefficients are just the squared magnitude of $r_s$ or $r_p$), and $n$ and $n_0$ are the refractive indices of two mediums, and $\theta_i$ and $\theta_t$ are the  angle of incidence and the angle of reflection, respectively, which satisfy Snell's law $n_0\sin\theta_i=n\sin\theta_t$. In addition, there will be an additional phase between the s-polarization and the p-polarization light.The angle of incidence which depends on the structure of the telescope will affect the polarization. Considering the transmitting antenna is a double off-axis parabolic structure with a scanning head, and the detailed optical parameters are shown in Table~\ref{TAB:Zemax} (not including the scanning head). 
   
\begin{table}[ht!]
\centering
\caption{\textbf{Optical design parameters of the transmitting antenna.}}
\label{TAB:Zemax}
	 \begin{tabular}{lllllll}  
		\toprule   
		Surface & Radius (mm) & Thickness (mm) & Glass & Semi-diameter (mm) &  Conic\\
        \midrule  
        OBJ & Infinity & Infinity & -- & Infinity &  0\\
		STO & Infinity & 812.5 & -- & 65.000 &  0\\ 
        3 & -1625 & -780.0 & Mirror & 190.000 &  -1\\ 
        4 & -65 & 1155.0 & Mirror & 7.600 &  -1\\
        5 & -- & -150.0 & -- & 13.153 & --\\ 
        IMA & Infinity & -- & -- & 1.966 & 0\\
        \bottomrule
	\end{tabular}
\end{table}

   The paraboloids satisfy the equation $z=c~\sqrt[]{x^2+y^2}$. The angle of incidence will be very small as the parameters shown. In addition, the offset angles caused by the beam expansion are also very small. In this situation, the reflectances for the s-polarization and the p-polarization light are nearly the same with an ignorable additional phase. The scanning head consists of two plane mirrors, which the angle of incidence is near $45^\circ$\cite{Han:2018:point-ahead}. Different from the structure above, there will be a different reflectances between s-polarization and p-polarization light with an additional phase with  only one layer of metal coating.
   
   With multilayer coatings( 50 layers) of $SiO_2$ and $Ta_5O_2$ applied by the Optowide company. The reflectances of s-polarization and p-polarization light can reach $99.9908\%$ and $99.8168\%$ with an additional phase about $0.9996\pi$ at 780 nm (the wavelength of the signal light) for the plane mirrors. At the same time, the performance of these paraboloidal mirrors are even better.  Meanwhile, the average reflectance is about $97.96\%$ at 532 nm (the wavelength of the beacon laser), which is good enough for the fine tracking. Actually, it will be very fragile with more layers even though it can improve the property.
   
\subsection{Local polarization test of the transmitting antenna}
\begin{figure}[ht!]
\centering\includegraphics[width=12 cm]{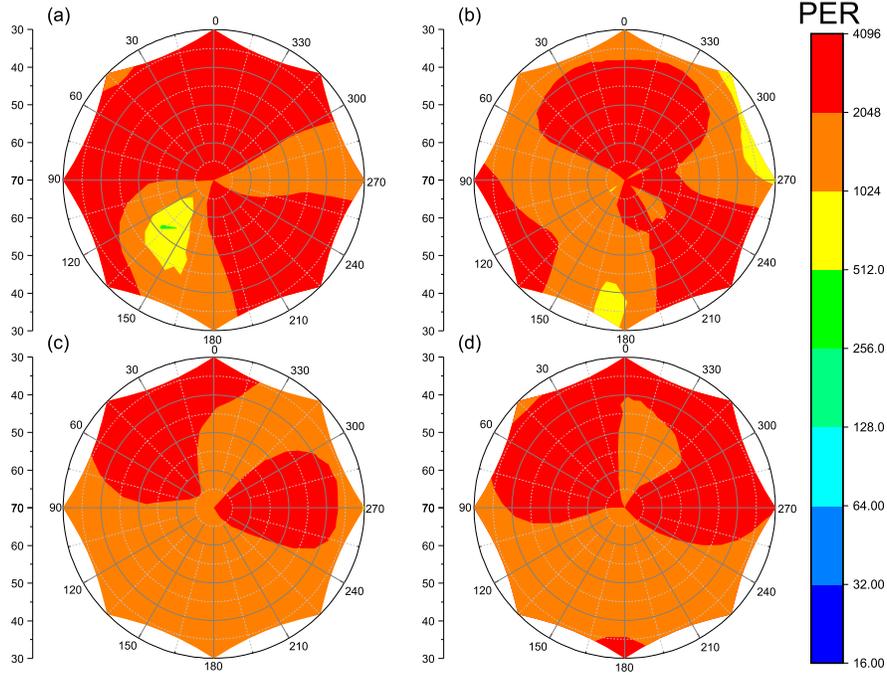}
\caption{\textbf{Local polarization test of the transmitting antenna.} Panels (a)-(d) show the PERs in different directions for $\left|H\right\rangle$, $\left|V\right\rangle$, $\left|+\right\rangle$ and $\left|-\right\rangle$, respectively.
}
\label{FIG:local}
\end{figure}

  To evaluate the performance of the coating system, a local test for polarization maintenance of the transmitting antenna is carried out. Since the passages of the satellite will change every day, the PERs must be tested for all directions of transmitting antenna. Without loss of generality, we selected several typical directions. The elevation angles of the transmitting antenna are set at $30^\circ$, $50^\circ$ and $70^\circ$, and the azimuth angles are set at $-180^\circ$, $-135^\circ$, $-90^\circ$, $-45^\circ$, $0^\circ$, $45^\circ$, $90^\circ$ and $135^\circ$. Four different polarizations of 780 nm lights  ($\left|H\right\rangle$, $\left|V\right\rangle$,$\left|+\right\rangle=(\left|H\right\rangle+\left|V\right\rangle)/~\sqrt[]{2}$, and $\left|-\right\rangle=(\left|H\right\rangle-\left|V\right\rangle)/~\sqrt[]{2}$) are successively transmitted from the collimator, where $\left|H\right\rangle$ and $\left|V\right\rangle$ represent the horizontal and vertical polarization states, respectively. The photons are collected using a large-diameter
collimator with a focal length of  1.5~$m$. Measurements of PERs are applied with a  convex lens, a polarizer and an optical power meter. The results of this local test are shown in Fig.~\ref{FIG:local}. The polarization fidelities for all possible incident directions exceed 0.9976 (PER = 445), with an average fidelity approaching 0.9988 (PER = 887). 
  
\section{Polarization compensation of the channel}
  The polarization compensation can be decomposed into two parts: (1) a constant polarization offset caused by fixed mirrors and fibers, and (2) time-dependent polarization changes caused by satellite motions. The constant polarization offset can be easily compensated by an half-wave plate (HWP) and two quarter-wave plates (QWPs)\cite{Englert:2001:universal_unitary}. Actually, the fiber connecting the transmitting antenna and the optical source must not be twisted while the transmitting antenna is rotating because that the polarization will vary with the fiber distortions which are not easy to measure and compensate simultaneously. 
  
  The transmitting antenna is therefor designed as a double off-axis reflective periscope structure with a scanning head\cite{Han:2018:point-ahead}, in which the fiber will remain stationary while the transmitting antenna is rotating. Other parts of the fiber are attached to the wall or the  table. A mirror can flip down to reflect the signal to a polarization detector. By adjusting the three plates, the PERs remain at better than 200:1 for more than 3 hours, even on a very windy ($20~m/s$) day. As the average passage of the satellite lasts about 530 seconds, thus the 3-hour stability of polarization compensation is enough to finish the experiment between the ground and the satellite.

\subsection{The compensation scheme for satellite motions}
  The relative motions between the satellite and the transmitting antenna changes the reference frame\cite{bonato:2006:sat_motion_polarization}, which can also influence the polarization. When the normal direction of the reference frame changes angle rotates an angle of $\theta$. Considering an observer who always measures along the propagation of light, the change of the direction of the normal angle of the reference will equal the change of the direction of the light. The polarization, therefore, changes along the direction, which is linear with the angle of direction. In our system, the direction of the scanning head will change while tracking the satellite, and so does the receiving telescope on the satellite. The directions of the propagation of light are determined by the transmitting antenna on the ground and the telescope on the satellite. These angles are easily obtained from the  two-line elements (TLE) data. The change can be dynamically compensated by a HWP automatically if the relationship is found. 
  
\begin{figure}[ht!]
\centering\includegraphics[width=12 cm]{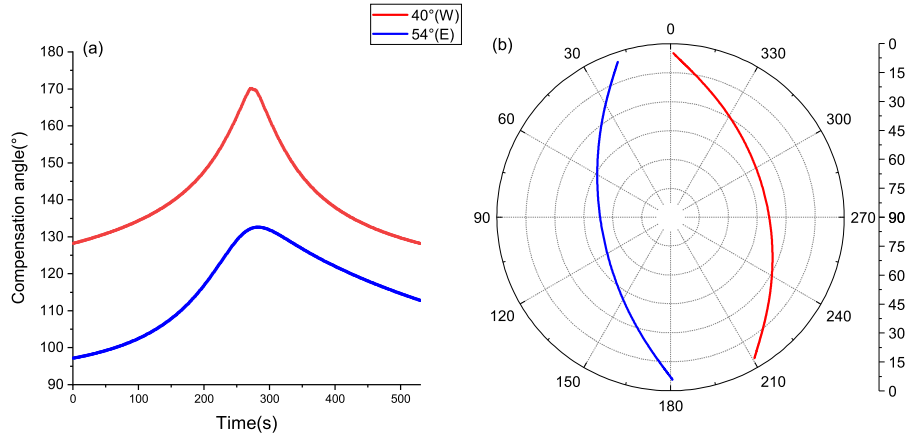}
\caption{
 \textbf{Compensation angles for two typical passages.}  (a) Compensation angles changing with time for two typical passages, which are shown in Panel (b). In the sky view (b),  $0^\circ$, $90^\circ$, $180^\circ$ and $270^\circ$ represent north, east, south and west, respectively in azimuth axis. On the polar axis shown on the right of Panel (b),  $0$ represents horizon and $90$ represents zenith.}                                                                                             
\label{FIG:compensation_angle} 
\end{figure}

  Initially, we must determine the zero point of the setup. A 780 nm $\left|H\right\rangle$ polarization light is transmitted from the collimator when the azimuth angle and the elevation angle of transmitting antenna is set at ($0^\circ$, $0^\circ$). When the automatic HWP rotates at $145.8^\circ$, the polarization becomes $\left|H\right\rangle$ again. When the azimuth angle and the elevation angle of transmitting antenna is at ($\theta$,$\varphi$), the polarization angle will increase $\theta +\varphi$. When the angle of the telescope in the satellite is $\beta$, the polarization angle will increase $\beta$. Because of multiple reflection, when the automatic HWP is at the angle $\alpha$, the polarization angle will decrease $\alpha/2$. Thus, as described above, the angle of the HWP is $145.8^\circ+(\theta+\varphi+\beta)/2$. With $(\theta, \varphi, \beta)$ from TLE data, the compensation angles against satellite motions can be easily calculated. Actually, the tracking errors are less than 10 $\mu rad$, which have negligible effects. The compensation angles are shown in Fig.~\ref{FIG:compensation_angle} for two typical passages. The passages are all from the north to the south at night. The change gradient of the compensation angels are less than $0.5^\circ/s$, which is much slower than the maximum rotational rate of the automatic HWP. The accuracy of the automatic HWP is about $0.01^\circ$, which will introduce an error of about $0.017\%$. Compared to the PER of the transmitting antenna, this can be ignored.
 
\subsection{Uplink polarization test}
To evaluate the performance of the polarization compensation algorithm, which gives the corrected compensation angles for the polarization change induced by the relative motion between the ground and the satellite, a polarization maintenance test is applied in the uplink channel. The photons sent to the satellite are polarized at  $\left|H\right\rangle$, and the compensation is applied in the transmitting antenna by an electronically controlled half-wave plate, whose rotation angles come from the polarization compensation algorithm. The compensation angles for two typical passages are shown in Fig.~\ref{FIG:compensation_angle}.

 \begin{figure}[ht!]
\centering\includegraphics[width=8 cm]{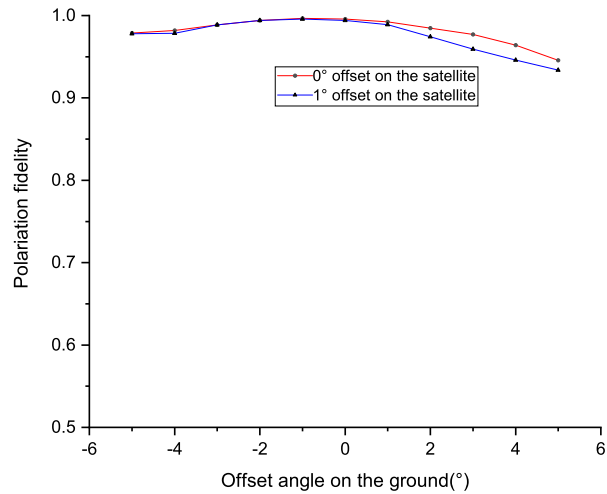}
\caption{\textbf{Polarization test from ground to satellite.} Offset angles are added both on the ground and satellite with every scanning point staying for 3 s, and the average fidelities of each scanning point are given.  
}
\label{FIG:gound-satellite}
\end{figure}

There are some offset angles both on the ground(from $-5^\circ$ to $5^\circ$ at $1^\circ$ intervals) and  satellite( $0^\circ$ and $-1^\circ$) added for saving time. As shown in Fig.~\ref{FIG:gound-satellite}, the fidelity is more than 0.995 when the offset is $(0^\circ,0^\circ)$, which is nearly the finest offset during the test. That means the compensation model against the relative motions is adequately within the margin of error. Some effects such as the temperature may influence accuracy of the fidelity.

\section{Ground to satellite entanglement distribution}

\begin{figure}[ht!]
\centering\includegraphics[width=12 cm]{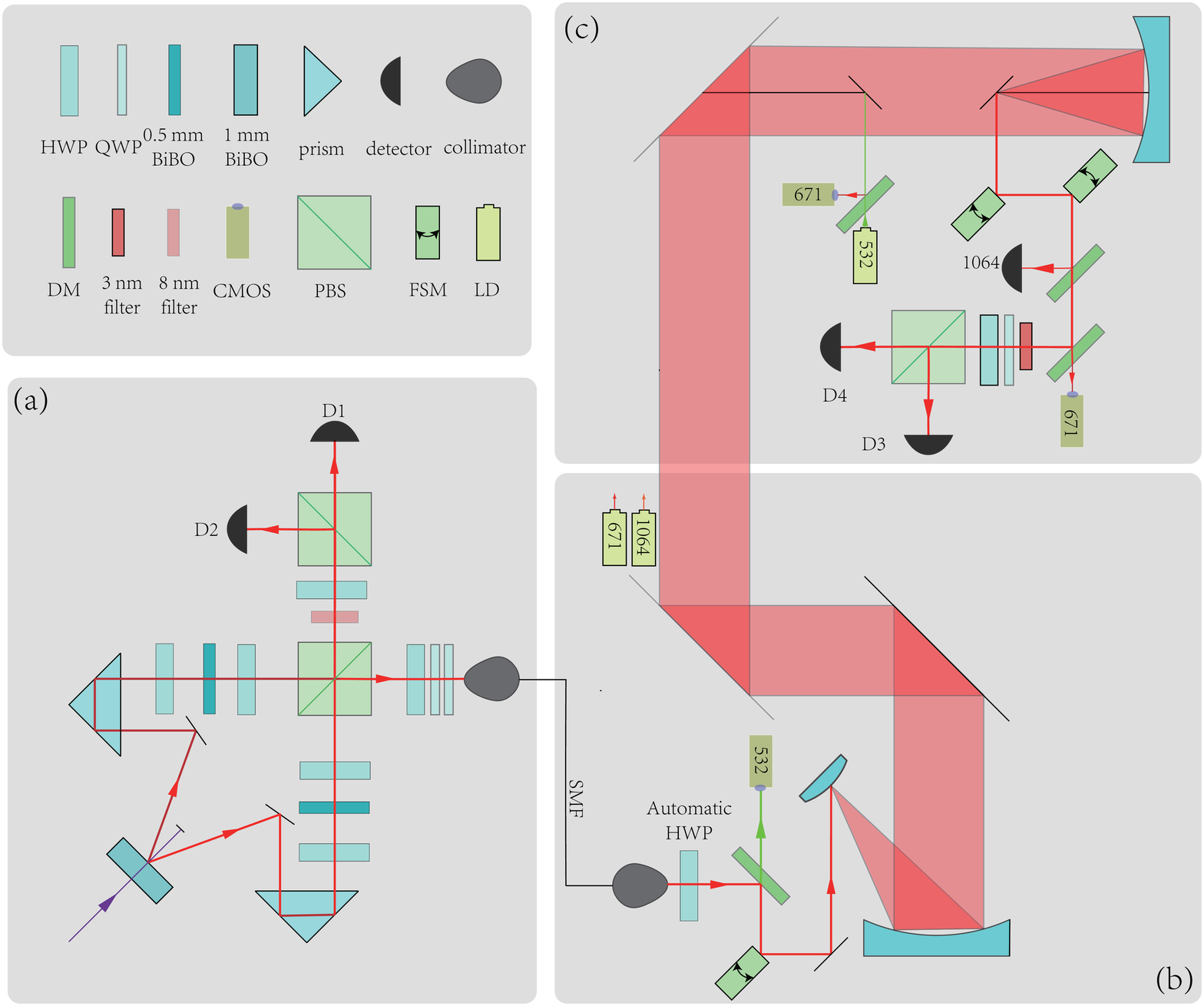}
\caption{\textbf{Experimental setup for ground-to-satellite Bell test.} (a) Entangled photon source. A 390 nm pulse with a repetition frequency of 80 MHz pumps a 1 mm  BiBO crystal. One photon is measured by the two detectors (D1 and D2) locally. Another photon is sent to the transmitting antenna via a 15-m-long single mode fiber (SMF). (b)Transmitting antenna. A 671 nm beacon laser diode (LD, power: 2 W, divergence angle: 1.2 mrad) and an 1064 nm synchronization LD (repetition rate: 10 kHz) are coaxial with the signal. Two cameras and a fast-steering mirror (FSM) are used for coarse tracking and fine tracking. An automatic half-wave plate (HWP) is used to compensate the polarization caused by satellite motions.  (c) Receiving system on the satellite. A 532 nm LD and two CMOS are used for tracking. The signal light is measured by D3 and D4. The synchronization light is measured by another detector(1064). Dichroic mirrors (DMs) are used to combine or split the signal light and the beacon laser. 
}
\label{FIG:setup}
\end{figure}

The experimental setup is shown as Fig.~\ref{FIG:setup}. The entangled sources and the transmitting antenna are located at the Ngari optical ground station ($32^\circ 19' 33.07''$ N, $80^\circ01'34.18''$ E, and altitude of 5,047 m). A 1.5 Watt 390-nm femtosecond pulse pumps a 1-mm bismuth borate (BiBO) to generate the down-converted photons. After compensation by the HWPs and BiBOs, the photons are entangled which can be expressed as:

 \begin{eqnarray}
\left|\Phi\right\rangle = \frac{1}{\sqrt[]{2}}(\left|HH\right\rangle+\left|VV\right\rangle)
\end{eqnarray}

The brightness of the entangled sources is about $1\times 10^6$ pairs/s with a fidelity of $93.29\%$. One photon is sent to satellite via a 15-m-length single mode fiber (SMF) and the transmitting antenna. Before the experiment, an HWP and two QWPs are used to compensate the polarization against the SMF with the fidelity more than 0.995. The other photon is measured locally by detector 1 and detector 2 (D1 and D2). The basis can be chosen by rotating the HWP and QWP before detecting both on the ground and on the satellite. All the photon events are recorded into time-to-digital converters (TDC). The uplink loss is approximately 41-52 dB. We use the CHSH-type Bell inequality \cite{CHSH_Bell_69} shown below:

   \begin{eqnarray}
   S=\left|E(\phi_1,\phi_2)-E(\phi_1,\phi_2')+E(\phi_1',\phi_2)+E(\phi_1',\phi_2') \right|
   \end{eqnarray}
   
Where $\phi_1$ and $\phi_1'$ are two measurement angels for photon 1 and $\phi_2$ and $\phi_2'$ are similar angles for photon 2. $E(\phi_1,\phi_2)$ represents the joint correlation which is defined as follows:
   \begin{eqnarray}
 E(\phi_1,\phi_2) = \frac{C(\phi_1,\phi_2)+C(\phi_1^{\bot},\phi_2^{\bot})-C(\phi_1,\phi_2^{\bot})-C(\phi_1^{\bot},\phi_2)}{C(\phi_1,\phi_2)+C(\phi_1^{\bot},\phi_2^{\bot})+C(\phi_1,\phi_2^{\bot})+C(\phi_1^{\bot},\phi_2)}
   \end{eqnarray}
  
\begin{figure}[ht!]
\centering\includegraphics[width=9 cm]{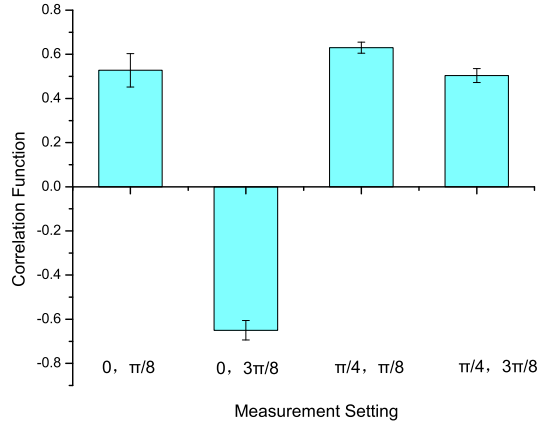}
\caption{\textbf{Experimental result for ground-to-satellite Bell test.} Four measurement settings $(0,\pi/8)$, $(0,3\pi/8)$, $(\pi/4,\pi/8)$, $(\pi/4,3\pi/8)$ are used. The error bars represent one standard deviation, calculated from the propagated Poissonian counting statistics of the events.
}
\label{FIG:result}
\end{figure}

 Where $\phi_1^{\bot}=\phi_1+\pi/2$, and $C(\phi_1,\phi_2)$ are the coincidence counts while photon 1 is measured at $\phi_1$ and photon 2 is measured at $\phi_2$. In our experiment, photon 1 are measured on the satellite at the angles $0$ or $\pi/4$ during one passage respectively. Photon 2 is measured on the ground at the angles $\pi/8$ or $3\pi/8$. We collect 2138 coincidence events with $S=2.312 \pm 0.096$. The detailed results of four settings are shown in Fig. \ref{FIG:result}. In addition to the downlink experiment\cite{Yin:SatEPR:2017}, it is another evidence for the non-local feature of entanglement on a large scale.
 
\section*{Conclusion}
In this paper, we design a transmitting antenna with polarization maintenance and achieve a polarization compensation scheme for satellite motions. The polarization fidelity is more than 0.995. This system can also be used for satellite-based classical communication\cite{toyoshima:2008:laser_communication} with polarization multiplexed. Furthermore, we distribute entangled photons from the ground to satellite with $S=2.312 \pm 0.096$. This result confirms the survival of entanglement in an uplink channel. Using a random basis in future experiments, we can build an uplink entanglement-based quantum key distribution, which is complementary to downlink QKD\cite{Yin:2017:satellite_QKD} for global quantum network.  

\section*{Funding}
This work was supported by the Strategic Priority Research Program on Space Science, the Chinese Academy of Sciences, National Natural Science Foundation of China Grants U1738203 and  U1738204, and Shanghai Sailing Program (No.18YF1425100 and No.17YF1420800).

\section*{Acknowledgments}
We thank many colleagues at Nanjing Astronomical Instruments Company Limited, the National Space Science Center, National Astronomical Observatories and Xi'an Satellite Control Centre and Optowide Company.

\section*{Disclosures}
The authors declare no conflicts of interest.


\bibliographystyle{osajnl}

\end{document}